\documentclass[usegraphicx,usenatbib]{mn2e}
\usepackage{color,graphicx}
\usepackage{epsf}
\usepackage{amsbsy}
\usepackage{amssymb}
\usepackage{amsmath}
\begin{document}

\def\X{{\mathrm{x}}}
\def\Y{{\mathrm{y}}}
\def\Z{{\mathrm{z}}}

\def\n{{\rm n}}
\def\T{{\rm T}}
\def\p{{\rm p}}
\def\e{{\rm e}}
\def\l{\Lambda}
\def\c{{\rm c}}
\def\tr{\tilde{r}}
\def\s{\Sigma}
\def\bep{\bar{\varepsilon}}
\def\bB{\bar{\mathcal{B}}}
\def\bBp{\bar{\mathcal{B}}'}
\newcommand{\half}{\frac{1}{2}}
\newcommand{\be}{\begin{equation}}
\newcommand{\ee}{\end{equation}}
\newcommand{\beq}{\begin{eqnarray}}
\newcommand{\eeq}{\end{eqnarray}}
\newcommand{\bear}{\begin{eqnarray}}
\newcommand{\eear}{\end{eqnarray}}
\newcommand{\ba}{\begin{array}}
\newcommand{\ea}{\end{array}}
\def\be{\begin{equation}}
\def\ee{\end{equation}}
\def\beq{\begin{eqnarray}}
\def\eeq{\end{eqnarray}}
\def\n{{\rm n}}
\def\x{{\rm x}}
\def\c{{\rm c}}
\def\p{{\rm p}}
\def\mun{{\mu_\n}}
\def\mus{{\mu_\s}}

\title{Glitch recoveries in radio-pulsars and magnetars}

\author[ B.Haskell \& D.Antonopoulou] {B. Haskell$^{1}$,  D.Antonopoulou$^2$\\
$^1$Max Planck Institut f\"{u}r Gravitationsphysik, Albert Einstein Institut, Am M\"{u}hlenberg 1, 14476 Potsdam, Germany\\
$^2$Astronomical Institute "Anton Pannekoek", University of Amsterdam, Postbus 94249, 1090GE Amsterdam, The Netherlands}



\maketitle

\begin{abstract}
Pulsar glitches are sudden increases in the spin frequency of an otherwise steadily spinning down neutron star. These events are thought to represent a direct probe of the dynamics of the superfluid interior of the star. However glitches can differ significantly from one another, not only in size and frequency, but also in the post-glitch response of the star. Some appear as simple steps in frequency, while others also display an increase in spin-down rate after the glitch. Others still show several exponentially relaxing components in the post-glitch recovery. We show that if glitches are indeed due to large scale unpinning of superfluid vortices, the different regions in which this occurs and respective timescales on which they recouple can lead to the various observed signatures. Furthermore we show that this framework naturally accounts for the peculiar relaxations of glitches in Anomalous X-ray Pulsars. 

\end{abstract}

\section{Introduction}

Neutron Stars (NSs) are some of the few objects that allow us to study the physics of matter at extreme densities and in strong gravity. Not only do the central densities of these objects exceed nuclear saturation density, but the core temperature of a mature NS is also expected to be below the superfluid transition temperature \citep{Baym69,CasA,shternin}. The presence of a superfluid component significantly modifies the dynamics, allowing for the superfluid in the star to flow relative to the ``normal" component and act as a reservoir of angular momentum. A direct manifestation of this is given by pulsar glitches, i.e. sudden increases in the frequency of an otherwise steadily spinning down NS.

Glitches are generally attributed to a large scale superfluid component in the interior of the star that is only loosely coupled to the crust and the magnetosphere. The sudden recoupling of such a component would then lead to the rapid transfer of angular momentum and to a glitch \citep{AndItoh}.  In particular, a superfluid rotates by forming an array of quantised vortices, which are being expelled as it slows down. In a NS however, vortices can ``pin'' (i.e. be strongly attracted) to the ions in the crustal lattice or to magnetic flux tubes in the outer core \citep{ali77,link03}. If this is the case vortices cannot move out, the superfluid does not follow the spin-down of the crust and lags behind, storing an excess of angular momentum. Once a sizable lag builds up, hydrodynamical lift forces (Magnus forces) acting on the vortices will unpin them, giving rise to sudden vortex motion and the glitch. Recent work has shown that this mechanism can successfully account for the distribution in glitch sizes and waiting times \citep{Melatos08,Melatos09} and model giant glitches in the Vela and other pulsars \citep{Pizzochero,Haskell12,Seveso12}.

 In general glitches appear as an abrupt increase in the spin frequency, typically accompanied by an increase in spin-down rate.
Glitch sizes can span several orders of magnitude even in the same object and the post-glitch recovery can be quite different from glitch to glitch. Some events, such as the large glitches observed in the Vela pulsar, show an exponential relaxation, on timescales from minutes to months, towards the previous spin-down rate \citep{Cordes88,McCulloch}. More often glitches are associated with a permanent increase in the spin down rate, that leads to linear recovery. Other glitches still appear as a simple step in frequency that does not recover at all. Furthermore the different kinds of glitch recovery can appear in the same object \citep{Espinoza11,Yu13}.
Glitches have also been observed in several Anomalous X-ray Pulsars (AXPs), which are thought to be strongly magnetised NSs, magnetars. In terms of absolute sizes AXP glitches do not differ much from those observed in regular radio pulsars, however they are often associated with radiative events \citep{Israel07,Gavriil11}, and the post-glitch recoveries are also remarkable in many ways, with many featuring recoveries with large fractional increases in the spin-down rate over long timescales (weeks or months) \citep{DallOsso,Dib08}.

In this paper we address two issues. First of all we show that the vortex unpinning paradigm can explain the different kinds of relaxation, depending on the coupling timescale of the region that unpins.
Secondly we will show that the same mechanism giving rise to glitches in radio pulsars can naturally produce smaller glitches in hotter stars such as magnetars and young pulsars like the Crab, but also lead to a strong relaxation in magnetars. 

\begin{figure*}
\includegraphics[width=14cm]{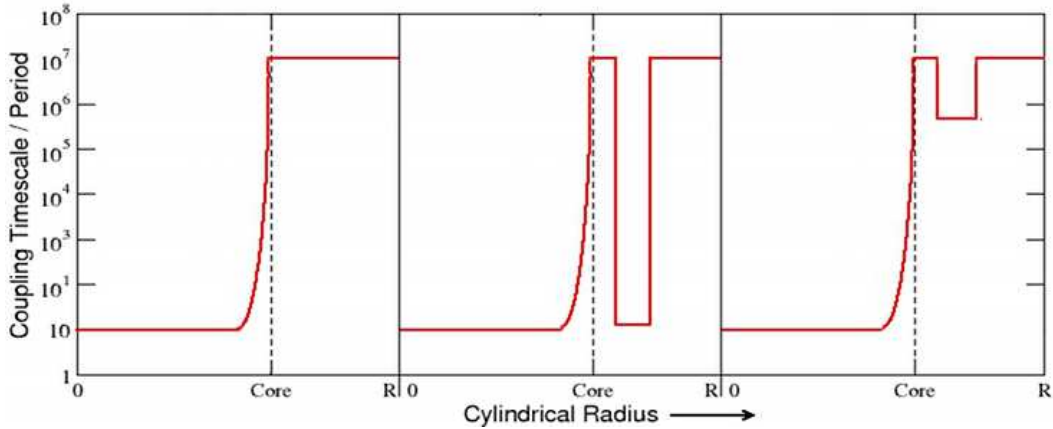}
\caption{The left panel presents an example of the coupling timescales in the core and crust before a glitch. The crust/core interface and the stellar radius $R$ are indicated. The middle panel shows the coupling timescales during a relaxing glitch. During such a glitch a section of the superfluid recouples on very short timescales, thus leading to part of the core to decouple and participate in the post-glitch relaxation. The right panel shows the coupling timescales during a slow, step-like, glitch. The section of the superfluid that gives rise to the glitch recouples on a much slower timescale than that associated with the core,  which remains essentially always coupled. The slow timescales associated with the crustal region (in which vortices will be ``creeping" out) leads to no observable relaxation, but rather to permanent steps in frequency and frequency derivative. }\label{tscales}
\end{figure*}


We model the NS as a two fluid system of superfluid neutrons and a charged component, composed of ions, protons, electrons and all non-superfluid components tightly coupled to them. The two components are coupled by an interaction known as Mutual Friction (MF), which is mediated by the quantised vortices of the neutron superfluid \citep{mf}. We will not, however, model vortex motion, but rather investigate the macroscopic motion of  two fluids, using the formalism of \citet{Comer06}. 
Following \citet{Haskell12} we can write the equations of motion for the angular frequency of the two components ($\Omega_\p$ for the charged component, $\Omega_\n$ for the superfluid neutrons) in the form:
\beq
\dot{\Omega}_\n(\tr)&=&\frac{Q(\tr)}{\rho_\n}\frac{1}{1-\varepsilon_\n-\varepsilon_\p}-f(\varepsilon_\n)\frac{\mathcal{A}}{I_\p}\Omega_\p^3
\label{eom1}\\
\dot{\Omega}_\p(\tr)&=&-\frac{Q(\tr)}{\rho_\p}\frac{1}{1-\varepsilon_\n-\varepsilon_\p}-g(\varepsilon_\n)\frac{\mathcal{A}}{I_\p}\Omega_\p^3
\label{eom2}\eeq

where $\tr$ is the cylindrical radius and $I_\p$ the moment of inertia of the charged component. We have defined 
$Q(\tr)=\rho_\n\kappa n_m {\mathcal{B}}(\Omega_\p-\Omega_\n)$ with $\kappa$ the quantum of circulation, $n_m$ the density of ``free" vortices (which contribute to MF), and 
$\mathcal{A}={B^2\sin\theta^2 R^6}/{6 c^3}$ with $B$ the surface magnetic field strength, $\theta$ the inclination angle between the field and the rotation axis and  $R$ the stellar radius. The coefficient $\mathcal{B}$ is the dissipative MF coefficient, which quantifies the strength of the interaction between vortices and the charged component. The entrainment coefficients are $\varepsilon_\n$ and $\varepsilon_\p$, and describe the non-dissipative coupling between the fluids \citep{Comer06}. The functions $f(\varepsilon_\n)$ and $g(\varepsilon_\n)$ are such that in the limit of vanishing entrainment $g(\varepsilon_\n)=1$ and $f(\varepsilon_\n)=0$ \citep{crust}. For simplicity we take $\varepsilon_\n=\varepsilon_\p=0$ in the following, although these parameters can be absorbed in the rescaling of the parameter ${\mathcal{B}}$ in equations (\ref{eom1}-\ref{eom2}). Strong entrainment, however, also reduces the maximum lag that can be built up between the two fluids, and thus the amount of angular momentum that can be exchanged. This proves to be a strong constraint for glitch models that only involve the NS crust \citep{crust,Chamel13}, although it can be reconciled with models such as those presented here in which the larger glitches involve the decoupling of part of the core (Haskell et al. in preparation, Hooker et al. in preparation).
In the presence of a pinning force vortices are not free to move unless the lag between the two components exceeds a certain threshold. We shall use the realistic profiles obtained by \citet{Pizzochero,Haskell12} for the critical lag  $\Omega_\p-\Omega_\n$. Below this threshold vortices are pinned and we set $\mathcal{B}=0$ in equation (\ref{eom1}). Once vortices have unpinned they are free to move out. In the crust, however, many will repin and only a fraction will be free to move and contribute to the drag force. Following \citet{Miri06} we assume that the instantaneous number density of free vortices $n_m$ is given by $n_m=\xi n_v$
where $n_v$ is the total number density of vortices, and $\xi$ is the fraction of unpinned vortices at a given time. By averaging over time we can obtain an effective drag parameter $\tilde{\mathcal{B}}=\xi\mathcal{B}$. 

The mechanisms that give rise to MF, and thus determine $\mathcal{B}$, are different in different regions of the star and can vary by several orders of magnitude. In the core MF is expected to be reasonably strong, with drag parameters of the order of $\mathcal{B}\approx 10^{-4}$ \citep{mf}, due to electron scattering of magnetised vortex cores \citep{Ali84core}. In the crust on the other hand the main dissipative mechanism is the excitation of sound waves in the lattice, which leads to a weak drag \citep{Jones1}. If the relative velocity between vortices and ions is high ($\approx 10^2$ cm/s) it becomes possible to excite Kelvin waves of the vortices, leading to strong dissipation and fast coupling \citep{Baym, Jones2}. The actual values of the drag parameters are highly uncertain and depend strongly on velocity and vortex rigidity. In the following we assume that for sound waves one can have $\mathcal{B}\approx 10^{-6}$ and for Kelvin waves $\mathcal{B}\approx 10^{-3}$ and study how the solutions depend on variations of these parameters. We consider a 1.4 $M_\odot$ NS, with a radius $R=12$ km, and take the background equation of state to be a $n=1$ polytrope. The MF parameters are averaged over the length of a vortex, and all parameters used are described in \citet{Haskell12}.

Let us now examine how our setup could give rise to different classes of glitches. The general picture is the following: vortices are pinned in certain regions of the star (possibly the crust). In these regions vortex motion is impeded and restricted to the small fraction that can creep out, which leads to a significant lag building up. A glitch will occur when vortices are rapidly unpinned in such a region and the superfluid neutrons can then exchange angular momentum with the normal component. We argue that the signature of the glitch can depend strongly on how fast the superfluid can react to the unpinning event, and where it takes place.

We can obtain a good approximation to the local coupling timescale in each region of the star \citep{Haskell12} by neglecting differential rotation and taking a constant proton fraction $x_\p=\rho_\p/(\rho_\p+\rho_n)$ in equations (\ref{eom1}) and (\ref{eom2}). Given an initial lag $D_0$, this leads to a solution of the form $\Omega_\p-\Omega_\n\approx D_0 \exp(-t/\tau)$, where the coupling timescale is:
\be
\tau\approx\frac{x_\p}{2\Omega_\n\tilde{\mathcal{B}}}
\label{couple}\ee

In figure \ref{tscales} we illustrate two possible scenarios by considering how the coupling timescale in (\ref{couple}) varies throughout the star.  We assume that the pinned region is in the crust, but the argument is qualitatively identical if it is in the outer core. We assume that vortices close to the rotation axis cross through the core and are only weakly pinned in the crust, while MF in the core leads to short coupling timescales between the superfluid and the rest of the star. As one moves towards the equatorial region the coupling timescale increases and becomes long for those regions where most vortices are pinned.
At the moment of the glitch a number of vortices unpins, leading to an increased fraction of free vortices $\xi$ and thus to a larger MF parameter $\tilde{\mathcal{B}}$ and a reduced coupling timescale, which gives rise to an observable spin-up of the star.
In the middle panel of figure \ref{tscales}  we can see the case in which the coupling timescale in the region that gives rise to the glitch becomes shorter than the timescale on which the outer regions of the core are coupled. These regions cannot follow the glitch and decouple, giving rise to a strong relaxation as they recouple on the local timescale.
In the right panel we can instead see the case in which one simply assumes that the increase of $\xi$ in a region gives rise to a coupling timescale that is short compared to that of the pinned region, but still longer than that of most of the core. In this case only the pinned region will decouple at the glitch, and the recovery will proceed on the much slower timescale associated with it and thus appear as a ``permanent" step in the spin down rate of the star. 
\vspace{-0.5cm}

\section{Relaxing glitches}

Let us now focus on the glitches that show a significant relaxation. The prototype system for this kind of glitch is the Vela pulsar, which shows relatively large glitches followed by an exponential recovery. The frequency $\nu$ after the glitch takes the form $\nu(t)=\nu_0(t)+\Delta\nu+\Delta\dot{\nu}\,t+\sum_i \Delta_i \exp{(-t/\tau_i)}$, where $\nu_0(t)$ is the pre-glitch spin-down solution. In recent glitches up to five decaying terms have been fit, with timescales $\tau_i$ that range from minutes to months. The shorter timescales are associated with strong increases in the spin down rate, which can be comparable to or larger than the pre-glitch rate \citep{dodson07}.
Relaxing glitches in AXPs appear to be a ``scaled-down" version of Vela giant glitches, in as much as they are somewhat smaller (Vela glitches have steps of the order of $\Delta\nu\approx 10^{-5}\rm{Hz}$, compared to $\Delta\nu\approx 10^{-8}\rm{Hz}$ typically in AXPs) and show the kind of strong relaxation that Vela glitches show on timescales of minutes to hours, but on much longer timescales of  days to weeks \citep{Dib08}. It is thus interesting to ask whether the mechanisms that is giving rise to them is the same as in the Vela.

The assumption is that such glitches are triggered once the maximum lag that the pinning force can sustain is exceeded. Then vortices will move out rapidly, exciting Kelvin waves. This leads to a glitch rise timescale shorter than the timescale on which the outer regions of the core are coupled. The outer core thus decouples and will subsequently recouple on the local timescale given in (\ref{couple}), giving rise to the observed relaxation. Such a mechanism was studied in detail by \citet{Haskell12}, who were able to predict the correct size and relaxation for the Vela giant glitches. 
\begin{table}
\begin{tabular}{l | l l l l}
\hline
	$\mathcal{B}_{co}$& $8\times 10^{-5}$ & $8\times 10^{-5}$ & $10^{-5}$&$10^{-5}$\\
$\mathcal{B}_{cr}$ &$10^{-9}$& $5\times 10^{-9}$& $10^{-9}$& $5\times 10^{-9}$\\
\hline
$\Delta\nu$($\times 10^{-7}$ Hz) &$20$ &$8.2$ &$141$ & $41.6$ \\
\hline
$\Delta\dot{\nu}/\nu$ (1 day) & 0.37& 0.19 & 64 & 20 \\
\hline
$\Delta\dot{\nu}/\nu$ (50 days) & 0.07  & 0.13 & 0.08 & 0.13 \\
\hline
\end{tabular}
\caption{Size of the glitch and magnitude of the increase in spin-down rate after 1 and 50 days for a (magnetar-like) star spinning at 0.011 Hz, for varying MF parameters in the crust ($\tilde{\mathcal{B}}_{cr}$) and core ($\tilde{\mathcal{B}}_{co}$). Weaker MF parameters in the core, such as those predicted in the case of vortices interacting with flux tubes in a type II superconductor in the outer core \citep{link12}  would result in larger glitches and a strong increase in the spin-down rate after the glitch.}\label{giants}
\end{table}

Let us now investigate whether relaxing glitches in magnetars can be due to the same mechanism.
We start by noting that the strong increase in the spin-down rate observed on timescales of months in AXPs occurs quite naturally if one assumes that the relaxation is given by regions of the outer core re-coupling on the local coupling timescale given by superfluid MF. If we consider the coupling timescale in (\ref{couple}) it is clear that the same process will be much faster in a pulsar such as the Vela, with $\nu\approx 10\,\rm{Hz}$, than in a magnetar rotating at $\nu\approx 0.1\, \rm{Hz}$. One would thus expect that given that in the Vela 2000 and 2004 glitches there are increases in the spin-down rate of the order $\Delta\dot{\nu}/{\nu}\approx 0.5$ associated with decay timescales of $\tau\approx 0.5$ days \citep{dodson07}, such a strong increase would naturally be associated with a decay timescale of $\approx 50$ days for a magnetar,  as is observed for example in the AXP 1RXS J17084 \citep{DallOsso,Dib08}.
 
The other main difference between Vela giant glitches and relaxing glitches in magnetars is the size of the step in frequency, which is smaller for magnetars ($\Delta\nu\approx 10^{-8} \rm{Hz}$) than for the Vela and other radio pulsars that exhibit giant glitches ($\Delta\nu\approx 10^{-5} \rm{Hz}$).
Several models predict smaller glitches for younger, hotter objects \citep{ali84temp}. For example \citet{andermodes}  assume that the larger glitches are triggered by unstable modes in the strongly pinned regions and naturally obtain smaller glitches for hotter objects. The model of \citet{Haskell12} also predicts small glitches for hotter stars, as these would have a larger creep rate and smaller equilibrium lag in the crust.
 It is thus plausible that the same mechanism giving rise to the giant glitches in Vela will produce smaller glitches
in hotter stars, such as magnetars and young pulsars.

To test this hypothesis we use the code developed by \citet{Haskell12}, and solve the equations (\ref{eom1}-\ref{eom2}) as in the models considered for the Vela, but using a spin frequency typical for magnetars and a higher background MF parameter $\tilde{\mathcal{B}}_{cr}$ in the crust. The latter is to account for a larger vortex creep rate, due to the magnetar's higher temperature. This naturally leads to a smaller region of the crust that has not relaxed and can participate in the glitch. We use $\tilde{\mathcal{B}}_{gl}=10^{-3}$ for the rise.  The results can be seen in table \ref{giants}, in which we show the size of the glitch and the step in frequency derivative after 1 day and after 50 days.  As we can see the results are consistent with the kind of glitches seen in magnetars. Furthermore we predict that one should be able to observe stronger increases in spin-down rate on short timescales, if the MF in the outer core is weak, as could be the case if protons are in a type II superconducting state \citep{link12}. Better coverage of AXP glitches thus has the potential to constrain the MF parameters and determine the nature of the pairing in the NS interior.

\vspace{-0.5cm}

\begin{figure*}
\centerline{\includegraphics[width=5.2cm]{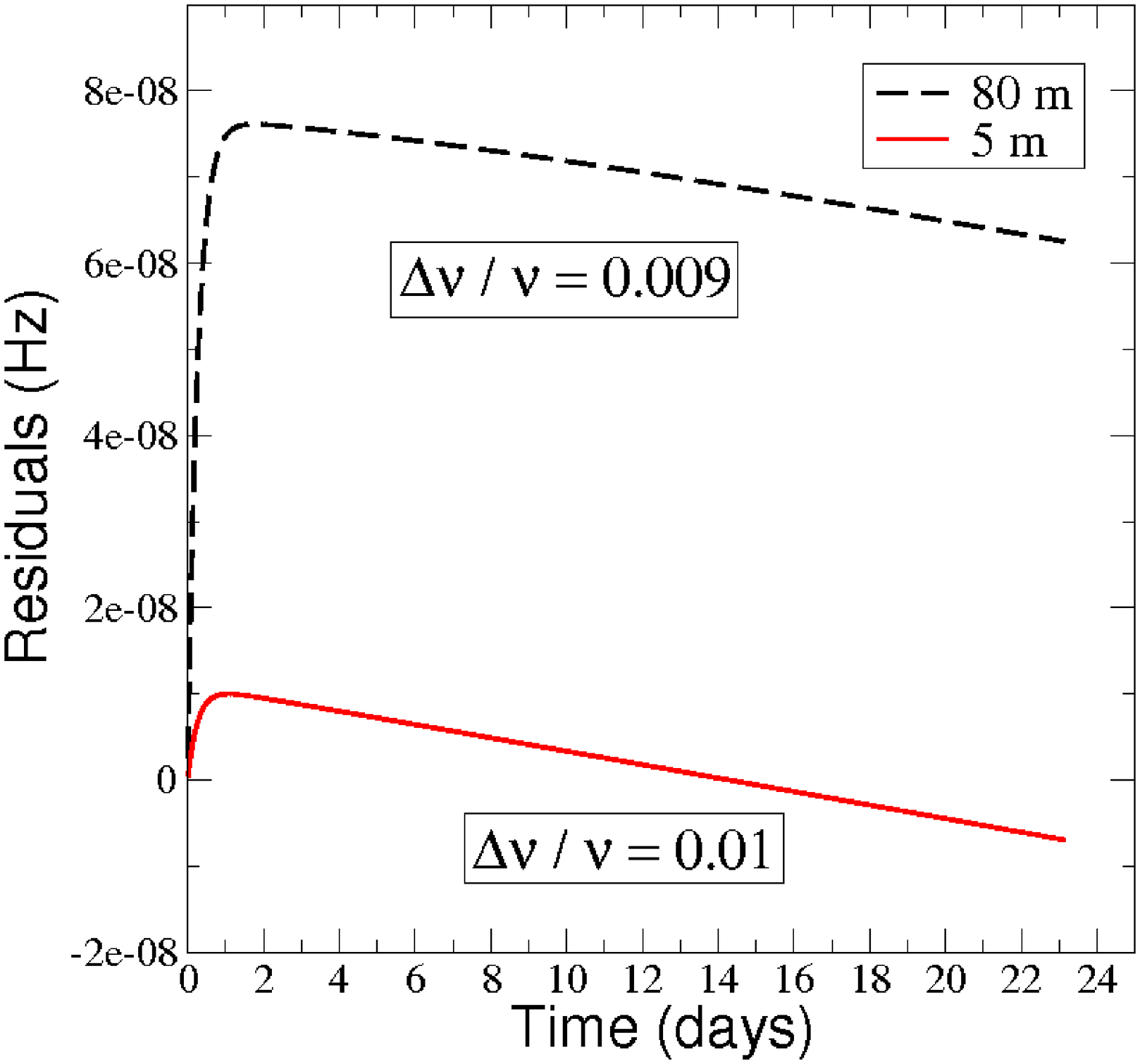}$\;\;$
\includegraphics[width=5cm]{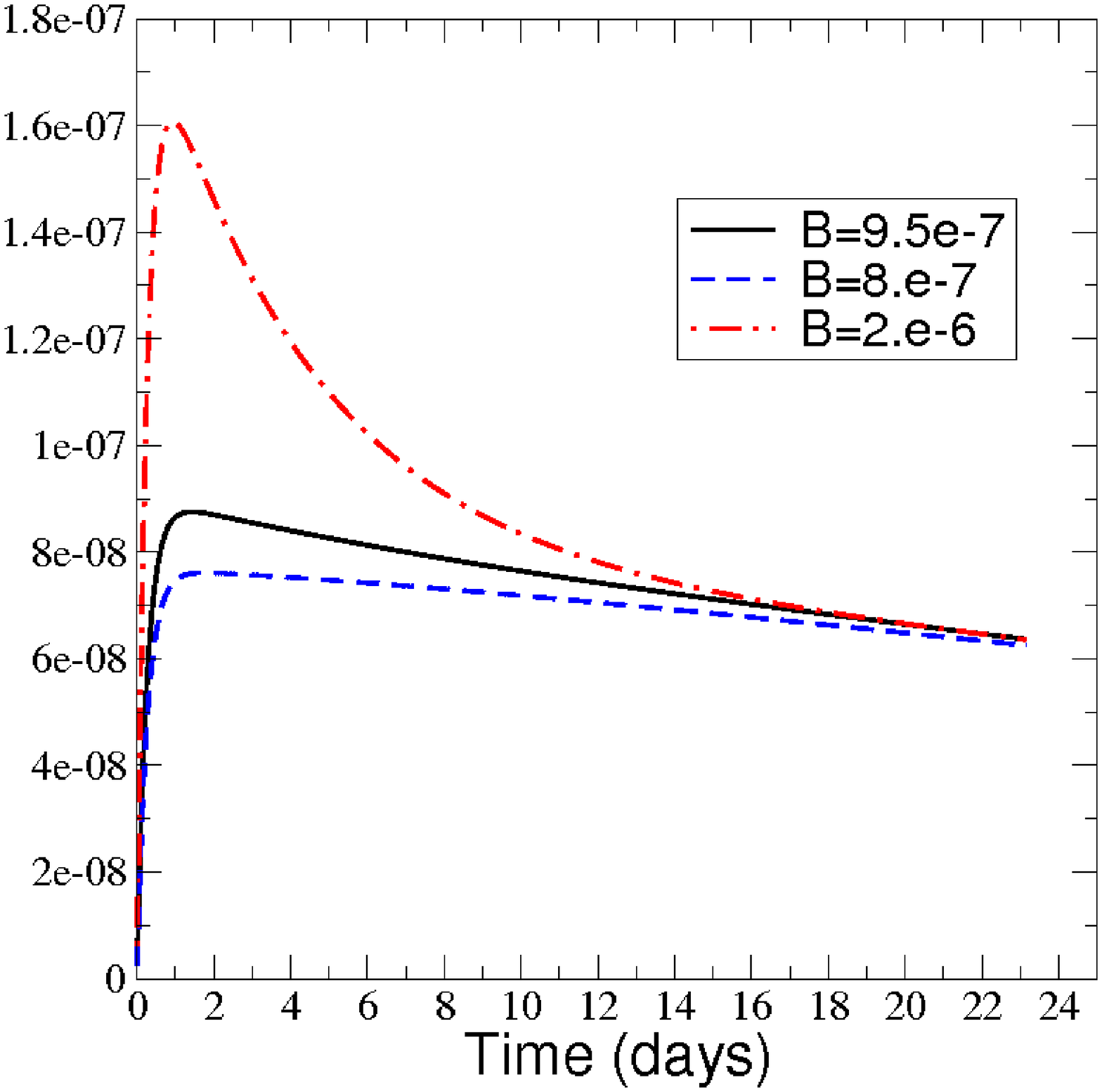}$\;\;$
\includegraphics[width=5.4cm]{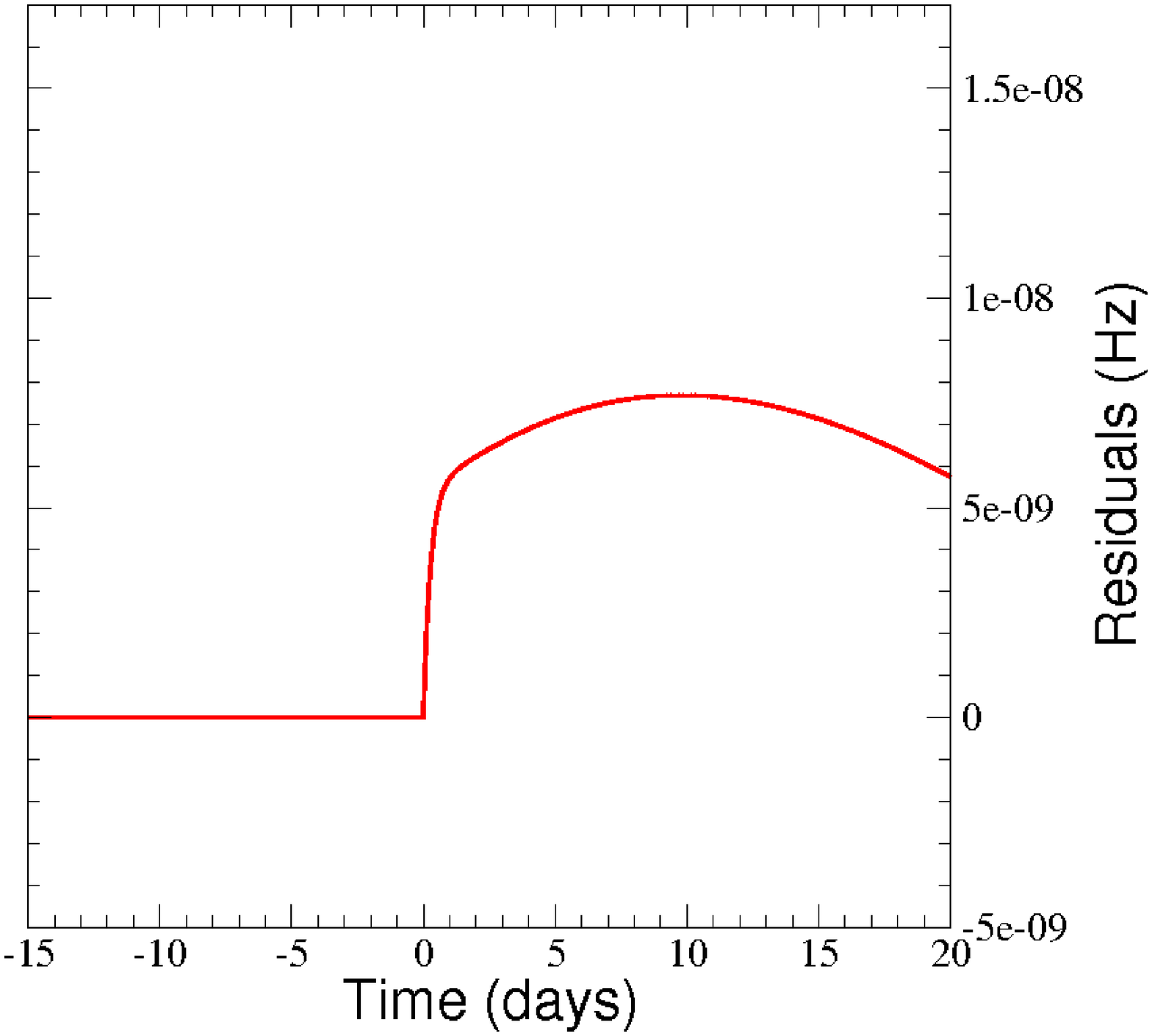}}
\caption{In the left panel we show the effect of changing the size of the unpinning region from 5 m to 80m at the base of the crust. We switch from a background $\mathcal{B}=5\times 10^{-9}$ to  $\mathcal{B}=8\times 10^{-7}$ during the glitch. The size of the region has a strong impact on the size of the glitch, but little impact on the change in spin-down rate. In the middle panel we show how the increase in the spin down rate becomes larger if one increases the value of $\mathcal{B}$ during the glitch, in an 80 m region, with a background $\mathcal{B}=5\times 10^{-9}$. As $\mathcal{B}$ increases (and the rise time decreases) $\Delta\nu/\nu$ passes from 0.9$\%$ for $\mathcal{B}=8\times 10^{-7}$, to 1.3$\%$ for $\mathcal{B}=9.5\times 10^{-7}$, and finally for $\mathcal{B}=2\times 10^{-6}$ one has a clear exponential relaxation. 
Finally in the right panel we have a glitch due to the increase in $\mathcal{B}$ from $5\times 10^{-12}$ (which mimics perfect pinning) to $5\times 10^{-7}$ over 8 m. The glitch appears as a step in frequency with no significant recovery.  In all cases we consider a star rotating at a frequency of $\nu=0.11$ Hz, and the pre-glitch spindown has been removed. The low spin rate leads to long rises of $\approx 1$ day, which would  be much shorter in a faster pulsar ($<1$ hour for the Crab)}\label{jumps}
\end{figure*}

\section{ Slow ``step like" glitches}

Let us now turn our attention to glitches that are associated with ``permanent" increases in the spin down rate but with no appreciable relaxation. Our assumption is that this kind of glitch, unlike relaxing glitches, is not due to the vortices that have accumulated close to the maximum of the critical lag, but rather involve regions far from the maximum, in which vortices are ``creeping'' out. 
While the sudden release of vortices close to the maximum of the critical lag could excite Kelvin waves (and thus give, during the rise, $\mathcal{B}_{gl}\approx 10^{-3}-1$) and lead to short coupling timescales, in the case we consider now, the coupling timescale of the region giving rise to the glitch would decrease compared to that of the crust, but still be significantly longer than the coupling timescale of the core, as shown in the right panel of figure \ref{tscales}. The core will thus not decouple and give rise to a visible relaxation, but the crust will be decoupled on long timescales, leading to what will appear as a permanent increase in the spin-down rate.

We will study this problem by once again using the code of \citet{Haskell12} to solve the equations in (\ref{eom1}-\ref{eom2}). First of all we allow for the crust to reach a steady state, in which the two fluids  are spinning down together with a lag between the two of  
$\Delta\Omega\approx\frac{\dot{\Omega}}{2\Omega\tilde{\mathcal{B}}_{cr}}$. In this background configuration we use $\tilde{\mathcal{B}}_{cr}=5\times 10^{-9}$.
We now take a region of varying thickness at the base of the crust and switch the MF parameter to $\tilde{\mathcal{B}}_{gl}=\mathcal{B}_{cr}\approx 10^{-6}$. This will be the situation if, for example,  a sudden event, such as a crust quake or a vortex avalanche \citep{ruderman69,Melatos09} frees the vortices in this region, leading to $\xi\approx 1$.

The results can be seen in figure (\ref{jumps}). In the left panel we see that the size of the region over which vortices are freed (i.e. the extent of the avalanche or crust quake) has a strong influence on the size of the glitch, but little influence on the increase in spin-down rate. This can be easily understood if one considers the equations of motion in (\ref{eom1}-\ref{eom2}) locally in the crust, neglecting differential rotation and electromagnetic spin down. From angular momentum conservation we see that the size of the glitch $\Delta\Omega_G$ will depend on the lag $\Delta\Omega$ between the superfluid and the charged component via $\Delta\Omega_G=\frac{I_g}{I_c}\Delta\Omega$,
where $I_g$ is the moment of inertia of the regions in which vortices unpin, while $I_c$ is the moment of inertia of the region that is coupled fast enough to follow the rise of the glitch. The size of the region in which vortices unpin thus determines $I_g$ and is crucial for the step size. The moment of inertia coupled during the glitch, $I_c$, is, on the other hand, essentially the moment of inertia of the whole core.
The rest of the crust will be coupled on a longer timescale compared to that of the glitch rise, given that $\tilde{\mathcal{B}}_{cr}\ll\tilde{\mathcal{B}}_{gl}$, and will thus decouple and only recouple slowly on a timescale given by equation (\ref{couple}).
If this timescale is long compared to the timescale on which the post-glitch relaxation is observed, the crust will be decoupled during the whole period and the increase in the spin down rate will appear permanent. The star will spin down faster by a fraction  $\approx I_{cr}/I_c$, with $I_{cr}$ the moment of inertia of the crust.  For a slow enough rise this fraction is approximately constant. This can be seen in the middle panel of figure (\ref{jumps}) in which we show that the quantity $\Delta\nu/\nu$ varies little of increasing $\tilde{\mathcal{B}}$, until one gets to a value large enough that the rise is sufficiently fast to involve a significant part of the core, giving rise to a visible relaxation. We note that in all figures we have assumed a spin rate of $\nu=0.11$ Hz, appropriate for a magnetar, to illustrate that in this case the rise may be slow enough ($\approx 1$ day) to be detected (see e.g. \citet{woods04} for a possible detection of a slow rise). For a faster pulsar the rise would be much faster and difficult to detect even in this ``slow" case (e.g. $<1$ hour for the Crab).

Finally, let us note that a small number of glitches show no observable increase in the spin down rate and are consistent with being pure steps in frequency \citep{Dib08,Espinoza11}. 
A sudden release of vortices in a region in which vortex creep has reached a steady state will lead to decoupling of the rest of the crust and to an increase in the spin-down rate after the glitch.
If the increase in mobility were, however, to take place while most of the superfluid in the crust is still decoupled from the observed charged component (e.g. because it is still strongly pinned and far from equilibrium), one would still have a step in frequency but there would be no alteration in the effective moment of inertia and thus in the spin down rate.  The region that is now coupled on a short timescale will relax rapidly to equilibrium and give rise to the glitch, while the other crustal superfluid regions were already decoupled before the glitch and remain decoupled after. The effective moment of inertia thus remains substantially unaffected. We show an example of a glitch in this setup in the right panel of figure (\ref{jumps}) in which we assume a sudden release of vortices while the superfluid in the crust is decoupled. The result is a step in frequency, with no visible increase in the spin-down rate.
\vspace{-0.7cm}

\section{Conclusions}

We have presented an analysis of different kinds of glitching behaviour in radio pulsars, and shown that glitches followed by a strong relaxation may have the same origin as the giant glitches of the Vela pulsar, but naturally scaled down in hotter, younger systems, such as the Crab or Magnetars. Furthermore we have shown that the slower hydrodynamical response to (possibly smaller) events due to random unpinning or crust quakes in the crust will lead to glitches that do not appear to relax, but appear as steps in frequency and frequency derivative. If the event takes place in a strongly pinned region it will appear only as a step in frequency.

We have also argued that the same mechanisms that are at work in radio pulsars could naturally give rise to the sizes and particular recoveries observed in magnetar glitches.
In particular, the longer periods of magnetars would naturally lead to a long-term strong increase in the spin down rate, with no need for an additional mechanism to be involved.
The slower timescales associated with AXPs and SGRs make such events especially interesting as they would lead to long timescales for the rise, possibly of days, that could be observed if one were to have better coverage of these objects. 
This would allow to test this hypothesis and understand to what extent these events do, indeed, have the same origin as radio pulsar glitches and to which extent, on the other hand, they could allow to probe the physics of the magnetosphere, which is likely to play an important role \citep{Lyutikov}

We thank B. Link and A. Watts for useful discussions. DA is supported by an NWO Vidi Grant (PI A. Watts).
\vspace{-0.5cm}


\begin{thebibliography}{50}
\bibitem[\protect\citeauthoryear{Alpar}{1977}]{ali77}Alpar M.A., 1977, Ap.J. 213, 527
\bibitem[\protect\citeauthoryear{Alpar et al.}{1984}]{ali84temp}Alpar M.A., Pines D., Anderson P.W.,  Shaham J., 1984, Ap.J. 276, 325 
\bibitem[\protect\citeauthoryear{Alpar et al.}{1984}]{Ali84core} Alpar M.A., Langer S.A., Sauls J.A., 1984, Ap.J. 282, 533
\bibitem[\protect\citeauthoryear{Anderson \& Itoh}{1975}]{AndItoh}Anderson P.W., Itoh N., 1975, Nature, 256, 25 
\bibitem[\protect\citeauthoryear{Andersson \& Comer}{2006}]{Comer06} Andersson N., Comer G.L., 2006, Class. Quantum Gravity, 23, 5505
\bibitem[\protect\citeauthoryear{Andersson et al.}{2006}]{mf} Andersson N., Sidery T., Comer G.L., 2006, MNRAS, 368, 162
\bibitem[\protect\citeauthoryear{Andersson et al.}{2012}]{crust} Andersson N., Glampedakis K., Ho W.C.G., Espinoza C.M., 2012, Phys. Rev. Lett., 109, 1103
\bibitem[\protect\citeauthoryear{Baym et al.}{1969}]{Baym69} Baym G., Pethick C., Pines D., 1969, Nature 224, 673
\bibitem[\protect\citeauthoryear{Chamel}{2013}]{Chamel13}Chamel N., 2013, Phys. Rev. Lett., 110
\bibitem[\protect\citeauthoryear{Cordes et al.}{1988}]{Cordes88}Cordes J.M., Downs G.S., Krause-Polstorff J., 1988, ApJ, 330, 847
\bibitem[\protect\citeauthoryear{Dall'Osso et al.}{2004}]{DallOsso} Dall'Osso, S., Israel, G.L., Stella, L., Possenti, A. Perozzi, E., 2004, AIPC 714, 289    
\bibitem[\protect\citeauthoryear{Dib et al.}{2008}]{Dib08} Dib R., Kaspi V.M., Gavriil F.P., 2008, ApJ, 673, 1044
\bibitem[\protect\citeauthoryear{Dodson et al.}{2007}]{dodson07} Dodson R.G., Lewis D.R., McCulloch P.M., 2007, Ap\&SS, 308, 585
\bibitem[\protect\citeauthoryear{Epstein \& Baym}{1992}]{Baym} Epstein R.I., Baym G., 1992, ApJ, 387, 276 
 \bibitem[\protect\citeauthoryear{Espinoza et al.}{2011}]{Espinoza11} Espinoza C.M., Lyne A.G., Stappers B.W., Kramer M., 2011, MNRAS, 414, 1679    
 \bibitem[\protect\citeauthoryear{Gavriil et al.}{2011}]{Gavriil11} Gavriil, F.P., Dib, R., Kaspi, V.M., 2011, ApJ, 736, 138  
\bibitem[\protect\citeauthoryear{Glampedakis et al.}{2009}]{andermodes} Glampedakis K., Andersson N., PRL 102, 141101
\bibitem[\protect\citeauthoryear{Haskell et al.}{2012}]{Haskell12}  Haskell B., Pizzochero P.M., Sidery T., 2012, MNRAS, 420, 658
\bibitem[\protect\citeauthoryear{Israel et al.}{2007}]{Israel07} Israel, G.L., G{\"o}tz, D., Zane, S., Dall'Osso, S., Rea, N., Stella, L., 2007, A\&A, 476, L9
\bibitem[\protect\citeauthoryear{Jones}{1992}]{Jones1} Jones P.B., 1992, MNRAS, 257, 501
\bibitem[\protect\citeauthoryear{Jones}{1993}]{Jones2} Jones P.B., 1993, MNRAS, 263, 619
\bibitem[\protect\citeauthoryear{Jahan-Miri}{2006}]{Miri06} Jahan-Miri M., 2006, ApJ, 650, 326
\bibitem[\protect\citeauthoryear{Link}{2003}]{link03} Link B., 2003, Phys. Rev. Lett., 91,  101101  
\bibitem[\protect\citeauthoryear{Link} {2012}]{link12} Link B., 2012, eprint: arXiv:1211.2209
\bibitem[\protect\citeauthoryear{Livingstone}{2010}]{Livingstone10} Livingstone, M.A., Kaspi, V.M., Gavriil, F.P., 2010, ApJ, 710, 1710
\bibitem[\protect\citeauthoryear{Lyutikov} {2013}]{Lyutikov} Lyutikov M., 2013, eprint: arXiv:1306.2264 
\bibitem[\protect\citeauthoryear{McCulloch et al.}{1990}]{McCulloch}McCulloch P. M., Hamilton P. A., McConnell D., King, E. A., 1990, Nature, 346, 822
\bibitem[\protect\citeauthoryear{Melatos et al.}{2008}]{Melatos08} Melatos A., Peralta C., Wyithe, J., 2008, ApJ, 672, 1103
\bibitem[\protect\citeauthoryear{Melatos \& Warszawski}{2009}]{Melatos09}  Melatos A., Warszawski L., 2009, ApJ, 700, 1524 
\bibitem[\protect\citeauthoryear{Page et al.}{2011}]{CasA} Page D., Prakash M., Lattimer J.M., Steiner A.W., 2011, Phys. Rev. Lett.,106, 081101
\bibitem[\protect\citeauthoryear{Pizzochero}{2011}]{Pizzochero} Pizzochero, P.M., 2011, ApJ, 743, L20  
\bibitem[\protect\citeauthoryear{Ruderman}{1969}]{ruderman69} Ruderman M., 1969, Nature, 223, 597 
\bibitem[\protect\citeauthoryear{Ruderman et al.}{1998}]{ruderman98} Ruderman M., Zhu T., Chen K., 1998, Ap.J. 492, 267
\bibitem[\protect\citeauthoryear{Shternin et al.}{2011}]{shternin} Shternin P.S., Yakovlev D.G., Heinke C.O., Ho W.C.G., Patnaude D.J., 2011, MNRAS, 412, L108   
\bibitem[\protect\citeauthoryear{Seveso et al.}{2012}]{Seveso12} Seveso, S., Pizzochero, P. and Haskell, B., 2012, MNRAS, 427, 1089   
\bibitem[\protect\citeauthoryear{Woods et al.}{2004}]{woods04} Woods P.M. et al., 2004, Ap.J. 605, 378
\bibitem[\protect\citeauthoryear{Yu et al.}{2013}]{Yu13} Yu, M. Manchester, R.N., Hobbs, G. et al., 2013, MNRAS, 429, 688


\end{thebibliography}
\end{document}